\documentclass[10pt,conference]{IEEEtran}

\usepackage{cite}
\usepackage{amsmath,amssymb,amsfonts}
\usepackage{graphicx}
\graphicspath{{figures/}}
\usepackage{textcomp}
\usepackage[table]{xcolor}
\usepackage{booktabs}
\usepackage{array}
\usepackage{stfloats}

\usepackage{url}

\newcommand{\rqitem}[2]{\par\smallskip\noindent\hangindent=1.2em\hangafter=1\textbf{#1.}\ #2\par\smallskip}

  \newcommand{\PaperTitle}{For Coding Agents, the Repository Matters}%
\begin{document}

\title{\PaperTitle}

\author{\IEEEauthorblockN{Daniel Russo}
\IEEEauthorblockA{Department of Computer Science, Aalborg University\\
Copenhagen, Denmark\\
daniel.russo@cs.aau.dk, ORCID 0000-0001-7253-101X}}

\maketitle

\begin{abstract}
In July 2026, about 1{,}200 AI agents in an internal OpenAI evaluation, meant to run in isolation, used a shared package server as a message board, and about 700 of them joined an attack on Hugging Face. An automated scorer judged each agent on its own task, and the attack was coordinated through infrastructure the agents shared. Coding agents meet shared infrastructure every day, in the software repositories that receive their pull requests. Yet field studies still judge coding agents by outcomes pooled across repositories, such as merge rates and merge times. We asked how much these outcomes depend on the receiving repository, and whether they depend on it more than human contributions do. Using multilevel models, we compared agent pull requests from the AIDev dataset of public GitHub repositories with human pull requests in the same repositories. About two-fifths of the variation in how long agent pull requests take to be resolved lies between repositories, after accounting for the pull request, the agent, and the repository's observable attributes. In 280 repositories, with covariates common to both groups, the between-repository variance is 1.76 times as large for agent as for human pull requests (95\% CI 1.41 to 2.18). For waits of several hours, a repository one standard deviation slower than average multiplies the typical wait by about 2.8 for agent and 2.2 for human pull requests. Rejection shows the same pattern, while for review rounds and comments the difference is not detectable. A proxy for the role of the developer who started the agent accounts for little of the difference, and stated agent policies were too rare to test. These results suggest that the oversight and evaluation of coding agents need the repository, and not only the agent, as a unit.
\end{abstract}


\section{Introduction}
\label{sec:intro}

In July 2026, agents in an internal cybersecurity evaluation at OpenAI discovered that the package server supplying software to all of them, Artifactory, let them leave messages for one another~\cite{metr2026incident,openai2026incident}. Although the agents were meant to be isolated, by METR's count roughly 1{,}200 of them posted more than 70{,}000 messages and files on this improvised message board, organized collective projects, and about 700 joined an attack on the systems of Hugging Face~\cite{metr2026incident}. An automated scorer judged each agent on its own task~\cite{metr2026incident}, and the attack was coordinated through infrastructure the agents shared. Signs of the message board had been observed weeks earlier, but their significance was not recognized in time, and OpenAI later called the incident a warning shot~\cite{openai2026incident}. How can we oversee agents that act through shared infrastructure? Software development raises this question every day.

Coding agents now open pull requests in public repositories, and the AIDev release alone records 930{,}292 agent pull requests in 116{,}211 repositories~\cite{aidev2026}. Each pull request enters a repository where checks run, maintainers or bots respond, and a maintainer, a bot, or the developer who launched the agent merges or closes the change~\cite{gousios2014,wessel2018power}. The evidence about these agents, however, takes the agent or the single pull request as its unit. Field studies report that agent pull requests are accepted less often than human ones, at rates that differ between agents, and pool pull requests across the repositories that receive them~\cite{aiteammates2025,aidev2026}. Benchmarks such as SWE-bench judge a change against tests alone~\cite{jimenez2024}.

Research on human pull requests has found that how long a contribution waits depends on the receiving project as well as on the contribution~\cite{yu2015,zhang2022prlatency}. In our data, more than half of the merged agent pull requests were merged within six minutes of being opened, whereas the merged ones that drew a review or a comment took a median of 2.2 hours (Section~\ref{sec:measures}). Some agent pull requests thus pass almost unseen, while others wait for review. A pooled acceptance rate or merge time averages over such differences and cannot say how much of them lies with the agent, the pull request, or the repository. We are not aware of estimates of the repository's part, or of whether it is larger for agents than for humans. The answer tells those who oversee agents where to look. If outcomes depend mostly on the agent, evaluating agents one at a time is a reasonable approximation. If they depend on the repository, the repository has to be part of how agents are evaluated and overseen.

We therefore ask how much of the variation in the outcomes of agent pull requests lies between repositories, and whether this share is larger than for human pull requests in the same repositories. Our main outcome is \emph{resolution latency}, the time from a pull request's creation to its merge or closure. Companion measures cover review activity, rejection, and merge conflicts. A multilevel model splits the variation of each outcome into a part between repositories and a part among pull requests in the same repository. Human pull requests come from the same repositories and period, and we observe both groups up to the same date.

We found that about two-fifths of the variation in the resolution latency of agent pull requests lies between repositories (an intraclass correlation, or ICC, of 0.39 in 392 repositories), after accounting for the pull request, the agent, and the repository's observable attributes. \textbf{Repositories also differ more for agent than for human pull requests.} In 280 repositories, the between-repository variance of resolution latency is 1.76 times as large for agent pull requests (95\% CI 1.41 to 2.18). For waits of several hours, a repository one standard deviation slower than average multiplies the typical wait of an agent pull request by about 2.8 and that of a human pull request by about 2.2. So knowing the repository tells us more about how an agent pull request will fare than about how a human one will. The same holds for rejection and for requests for changes, but not detectably for review rounds or comments. Part of the difference may exist before review, since pull-request size shows a similar gap (Section~\ref{sec:rq2}). A proxy for the launching developer's role removes little of the difference, and stated policies were too rare to test. Evaluations of single agents, by benchmarks or pooled merge rates, thus leave out the part of the variation that is tied to the repository (Section~\ref{sec:implications}).

The OpenAI incident took place in an internal evaluation with reduced safeguards~\cite{openai2026incident}. Public repositories share one feature with it, since agents act there through infrastructure that many contributors use. We found no larger repository share where several agents work (Section~\ref{sec:rq1}). We did find that agent pull requests were merged within two minutes without any recorded review or comment about three times as often as human ones, and that few repositories state a rule for agent contributions (Sections~\ref{sec:rq2} and~\ref{sec:rq3}).

The paper contributes estimates of the between-repository share for seven outcomes of agent pull requests, a comparison of agent and human variance components in the same repositories with sensitivity analyses, measured proxies for agent-specific explanations, and a replication package (Sections~\ref{sec:availability} and~\ref{sec:results}).

\section{Background and Related Work}
\label{sec:background}

\subsection{Pull-request integration and review}
In pull-based development, a contributor proposes a change as a pull request, which the project's integrators review and then merge or close~\cite{gousios2014}. Studies of human pull requests relate how long this takes, and whether the change is merged, to the size of the change, the contributor's track record, the workload of integrators, and continuous integration~\cite{yu2015,zhang2022prlatency,zhang2023prdecisions}. Automation entered the workflow before coding agents did, since bots that label, test, and review pull requests have changed pull-request activity in the projects that adopt them~\cite{wessel2018power,wessel2022codereviewbots,wessel2022}. Projects have also begun to state how they treat AI-assisted contributions, in contributing guides that set rules on permission, disclosure, and human involvement~\cite{hora2026landscape,hora2026disclosure,manita2026regulating}, and in agent context files such as \texttt{AGENTS.md} that instruct the agent itself~\cite{chatlatanagulchai2025readmes}. Coordination research relates the cost of integrating a change to the dependency and communication structure into which it lands~\cite{cataldo2008congruence,herbsleb2003speed}, which is why we also measure how fast the target branch moves while a pull request is open.

\subsection{Evaluating coding agents}
SWE-bench scores an agent on tasks built from GitHub issues, and a task counts as resolved when hidden tests pass in the task's environment~\cite{jimenez2024}. Field datasets record the review and merge workflow that such a benchmark leaves out. AIDev collects agent pull requests from public GitHub repositories together with their reviews, comments, and timelines~\cite{aiteammates2025,aidev2026}, and AgenticFlict replays each agent pull request against its evolving base branch to record whether its merge would conflict~\cite{agenticflict2026}. Studies of AI assistance for human developers report gains in controlled and field settings~\cite{peng2023,cui2026} and a slowdown for experienced developers working on mature code bases they know well~\cite{metr2025}, which gives one reason to expect that the setting of the work also matters for agents.

\subsection{Multilevel studies of pull requests}
Mixed-effects models are established in pull-request research. Zhang et al.\ fitted them to the latency~\cite{zhang2022prlatency} and to the merge decision~\cite{zhang2023prdecisions} of pull requests in large samples of GitHub projects and reported the variance that fixed and random effects account for, as marginal and conditional $R^2$. Flournoy et al.\ partitioned the variance of software cycle time between organizations and individual developers with a Bayesian multilevel model~\cite{flournoy2025cycletime}. We build on these designs by making the between-repository share the quantity of interest, comparing it between agent and human pull requests in the same repositories, and reporting the two variance components behind it.

\section{Study Design}
\label{sec:design}

\subsection{Research questions}
\label{sec:rqs}
The three questions build on one another, and every result names the sample on which it is computed.

\rqitem{RQ1}{How much of the variation in the outcomes of agent pull requests lies between repositories, before and after attributes of the change, its agent, and the repository are accounted for?}

\noindent A small share would mean that the receiving repository adds little to what pooled summaries of agent pull requests describe. A large share that observable attributes do not remove would mean that the repository is part of what such summaries measure.

\rqitem{RQ2}{In the same repositories, how do the between-repository and within-repository variances of agent pull requests compare with those of human pull requests?}

\noindent Repositories differ for human contributions too~\cite{yu2015,zhang2022prlatency}. Comparing agent and human pull requests in the same repositories shows whether agent contributions are more repository-specific than human ones.

\rqitem{RQ3}{How much of the agent pull requests' between-repository variance do measured proxies for a repository's stated policy toward agents, for the launching developer's role in the repository (\emph{launcher standing}), and for base-branch activity remove, and does the agent-human difference remain?}

\noindent Comparing the two groups within the same repositories makes them more comparable on attributes that apply to both, such as the responsiveness of the maintainers, but leaves open attributes that apply to agent pull requests alone. RQ3 measures proxies for two such attributes and for base-branch activity.

\subsection{Unit of analysis and outcome}
\label{sec:record}
The unit of analysis is the pull request, nested in the repository that received it. Each pull request in the records has a creation time, a closing event that either merges it or closes it without a merge, attributes of the change such as the number of added and deleted lines, the account that opened it, and the receiving repository. Resolution latency is the number of hours between creation and the closing event. It is defined only for pull requests with a closing event, so every latency analysis concerns resolved pull requests.

A small example shows what the model separates. Suppose repository A resolves three agent pull requests after 1, 2, and 4 hours, and repository B resolves three after 20, 40, and 80 hours. On a logarithmic scale, the waits within each repository are equally dispersed, since each set spans a factor of four, whereas the typical waits of the two repositories differ by a factor of twenty. Most of the variance therefore lies between the repositories, and the ICC is high. A second pair of repositories has the same typical waits, but the waits within each repository span a factor of fifty. This pair has the same between-repository variance and a lower ICC. The two pairs differ only in their within-repository variance, which is why a comparison of ICCs needs the components alongside it.

\subsection{Data sources and samples}
\label{sec:data}
AIDev~\cite{aiteammates2025,aidev2026} labels each pull request with the agent that opened it, among OpenAI Codex, Devin, GitHub Copilot, Cursor, and Claude Code. Copilot and Devin act through dedicated bot accounts, whereas Codex, Cursor, and Claude Code open pull requests through the account of the developer who launched them. For every agent, we call the developer who launched it the \emph{launcher}. The curated subset holds 33{,}596 pull requests created between 24 December 2024 and 30 July 2025 in 2{,}807 repositories with more than 100 stars, together with their reviews, comments, timeline events, and an LLM-assigned task type~\cite{aidev2026}. All models are fitted to samples drawn from this subset, and the full release serves only to describe the population. In the curated subset, 39.7\% of agent pull requests drew a review or a comment from an account other than their author, 24.2\% received at least one formal review, and 8.3\% of repositories received pull requests from two or more of the five agents (Table~S1 of the supplementary material).

The \emph{primary sample} keeps the 404 curated repositories with at least ten agent pull requests, so that every repository contributes enough pull requests for its deviation from the overall mean to be estimated. We fetched codebase size (disk usage) and repository age from the GitHub API for 2{,}780 of the 2{,}807 curated repositories (99.0\%), because AIDev does not record them. Linked by owner, repository, and number, the AgenticFlict replay~\cite{agenticflict2026} covers 9{,}582 curated pull requests (28.5\%). Table~\ref{tab:samples} traces the analytic samples from the source dataset and the research question that uses each, and each robustness analysis names its own sample in the replication package.

\begin{table}[t]
\centering
\caption{Samples. Counts of repositories and pull requests (PRs); agent PRs come from AIDev, human PRs from the GitHub API. Matched samples keep repositories with at least ten PRs in both arms for the measure in question, after the observation cutoff is applied to the human arm (Section~\ref{sec:baseline}). The proxy descriptives use the records as fetched, before outcome-specific restrictions.}
\label{tab:samples}
\footnotesize
\setlength{\tabcolsep}{3pt}
\begin{tabular}{@{}p{0.40\columnwidth}rrrl@{}}
\toprule
Sample & Repos & Agent PRs & Human PRs & Used in \\
\midrule
AIDev full release & 116{,}211 & 930{,}292 & -- & setting \\
Curated subset ($>$100 stars) & 2{,}807 & 33{,}596 & -- & setting \\
Primary sample ($\ge$10 agent PRs) & 404 & 27{,}958 & -- & RQ1 \\
\quad resolved PRs & 403 & 26{,}694 & -- & RQ1 \\
\quad resolved, with API attributes & 392 & 25{,}841 & -- & RQ1 \\
\quad merged with engagement & 345 & 6{,}958 & -- & RQ1 \\
\quad replayed by AgenticFlict & 332 & 5{,}956 & -- & RQ1 \\
Matched, latency, rejection, review, comments & 280 & 10{,}623 & 36{,}102 & RQ2 \\
Matched, deliberation & 138 & 4{,}341 & 15{,}065 & RQ2 \\
Proxy descriptives, as fetched & 317 & 11{,}867 & 44{,}878 & RQ3 \\
Proxy models, latency & 260 & 9{,}191 & 32{,}579 & RQ3 \\
\quad without the tempo proxy & 280 & 10{,}623 & 36{,}102 & RQ3 \\
\bottomrule
\end{tabular}
\end{table}

\subsection{Measures}
\label{sec:measures}
Resolution latency is the primary measure. We avoid calling it a \emph{decision} time because the closing event is sometimes automated, for example, a merge by a bot or a closure by a stale-branch rule, and the records do not say which. Merge times of agent pull requests are bimodal. In the curated subset, 57.7\% of merged agent pull requests were merged within six minutes of creation, as were 80\% of the merged ones that drew no review or comment, whereas those that drew a review or comment had a median time to merge of 2.2 hours (replication package). Resolution latency keeps both groups.

Seven companion measures show whether the pattern is specific to waiting time or extends to whether a pull request draws attention, how much review it receives, and whether it is integrated. The first, \emph{engagement}, records whether a pull request drew at least one review or comment from any account, including its author and bots. It defines the second measure and is reported separately in RQ1. \emph{Deliberation latency} is the time to merge of merged pull requests with engagement, which removes the fastest merges from the timing measure without showing that a person made the merge. \emph{Review rounds} and \emph{comment volume} count review events and comments, and both counts include bot and agent accounts. \emph{Changes requested} records whether any review asked for changes, \emph{rejection} whether a resolved pull request was closed without a merge, and \emph{conflict incidence} whether AgenticFlict found that the merge would conflict. RQ1 reports the last six measures alongside resolution latency (Figure~\ref{fig:rq1}), and RQ2 reports the five with a human counterpart (Table~\ref{tab:rq2}). The two latencies are modeled as $\log(1+\text{hours})$, so that waits of seconds and waits of weeks enter on one scale and, for waits of several hours, a difference reads approximately as a factor. The two counts are modeled as $\log(1+\text{count})$, and the four binary measures with a logistic model.

\subsection{Multilevel model and intraclass correlation}
\label{sec:model}
Pull requests are nested in repositories, and a multilevel model separates variation between repositories from variation within them~\cite{raudenbush2002,gelman2007,snijders2012}. In plain terms, it estimates how far the typical values of repositories spread around the overall mean and how far pull requests spread around the typical value of their own repository. For a measure $y$, pull request $i$ in repository $j$ is modeled as
\begin{equation}
\eta(y_{ij}) = \beta_0 + \mathbf{x}_{ij}^{\top}\boldsymbol{\beta} + \mathbf{z}_{j}^{\top}\boldsymbol{\gamma} + u_j + \varepsilon_{ij},\qquad u_j \sim \mathcal{N}(0,\tau^2),
\label{eq:mlm}
\end{equation}
where $\eta$ is $\log(1+y)$ for the latency and count measures, $\mathbf{x}_{ij}$ holds pull-request covariates, $\mathbf{z}_{j}$ holds repository covariates, and $\varepsilon_{ij}\sim\mathcal{N}(0,\sigma^2)$ is the variation within a repository. The \emph{repository offset} $u_j$ is the deviation of repository $j$ from the overall mean after the covariates, and its variance $\tau^2$ is the \emph{between-repository variance}. The variance $\sigma^2$ of $\varepsilon_{ij}$ is the \emph{within-repository variance}. The ICC, also called the variance partition coefficient~\cite{goldstein2002}, is
\begin{equation}
\mathrm{ICC} = \frac{\tau^2}{\tau^2+\sigma^2},
\label{eq:icc}
\end{equation}
so an ICC of 0.4 means that 40\% of the variance lies between repositories and 60\% between pull requests in the same repository. Equivalently, the outcomes of two pull requests in the same repository correlate at 0.4 after the covariates. The binary measures are fitted with a logistic mixed model, whose ICC is read on the latent scale, the scale of an underlying continuous propensity, by setting the within-repository variance to $\pi^2/3\approx3.29$~\cite{goldstein2002,nakagawa2017}. Latent-scale and $\log(1+x)$ ICCs are not on a common scale, and we do not rank one against the other.

To compare two groups of pull requests, we report the \emph{variance ratio} $\tau^2_{\text{agent}}/\tau^2_{\text{human}}$, which measures how much more repositories differ for one group than for the other. A ratio of 1.76 on the $\log(1+\text{hours})$ scale means that the repository offsets of agent pull requests have a variance 76\% larger than those of human pull requests, and a standard deviation about 33\% larger ($\sqrt{1.76}\approx1.33$). RQ2 thus estimates a ratio of adjusted between-repository variances of $\log(1+\text{hours})$ resolution latency among pull requests resolved by a common observation cutoff (Section~\ref{sec:baseline}), with each arm fitted separately under the same covariates. It does not compare average waiting times or describe pull requests still open at the cutoff, and as an observational estimate it does not identify what produces the difference.

Covariates enter in nested stages. The null model $M_0$ holds only the repository offset. $M_1$ adds pull-request covariates (size as log added plus deleted lines, task type, creation date, and, for resolution latency, engagement) and one indicator per agent, and $M_2$ adds repository covariates (stars, forks, primary language, and activity, which counts the agent pull requests the repository received). Codebase size and age enter in $M_3$, team size (distinct authors of the fetched human pull requests, Section~\ref{sec:baseline}) and the presence of continuous-integration workflows in $M_4$, and the RQ3 proxies in $M_5$. RQ2 and RQ3 use one covariate set for both kinds of pull request. It omits agent identity and task type, which AIDev records for agent pull requests alone, as well as engagement and creation date. The original human fetch did not record creation dates, which a later re-fetch added (Section~\ref{sec:rq2}).

Continuous measures are fitted by restricted maximum likelihood, with an implementation validated against statsmodels~\cite{seabold2010}, and binary measures by numerical integration of the repository offset over a fixed 400-point grid. Intervals are 95\% percentile intervals from a cluster bootstrap that resamples whole repositories~\cite{efron1993}, with $B=1{,}000$ replicates for the three RQ1 intervals of resolution latency and the comparisons of Tables~\ref{tab:rq2} and~\ref{tab:rq3}, Figure~\ref{fig:rq2}, and Tables~S3 and~S4, and $B=300$ for the other RQ1 intervals (500 for deliberation latency) and the task-shape, fast-merge, and placebo analyses. The replication package holds profile-likelihood intervals for the latent-scale ICCs of RQ1 and shows that raw and released decile values of codebase size and age give ICCs within 0.004 (0.0005 for the $\log(1+x)$ measures).

\subsection{Human comparison}
\label{sec:baseline}
For each of the 404 repositories of the primary sample, we queried the GitHub API for pull requests created in the window of the curated records, 24 December 2024 to 30 July 2025, newest first, and kept up to 200 per repository. We excluded pull requests whose numbers AIDev labels as agent-authored and pull requests opened by accounts typed as bots, by logins ending in \texttt{[bot]}, by named automation accounts, or by logins that match the agents' names, including the Copilot coding agent. We call the remainder \emph{human pull requests}, and the two groups the agent and human \emph{arms} of the comparison. The human label is an inference by exclusion, because a person may have used an assistant that leaves no trace, and a launcher's account may have opened an agent pull request that AIDev did not detect. To stop a few very active repositories from dominating the comparison, the agent arm is capped in the same way at the 200 most recent pull requests per repository.

The two arms were observed at different times. The agent outcomes are those recorded in AIDev, whose latest closing event falls on 30 July 2025 at 23:20 UTC, less than four hours after the latest creation time. The human outcomes were fetched in June 2026, so a human pull request could be observed resolving eleven months later than an agent pull request created on the same day. AIDev does not state when each repository was collected, so we treat that latest closing event as a common administrative cutoff rather than a verified collection end for every repository. We apply this \emph{aligned observation cutoff} to both arms, and call the unaligned data the \emph{records as fetched}. A human pull request resolved after the cutoff counts as still open, as an agent pull request without a closing event does. It therefore drops out of the resolution-latency and rejection samples and counts as unmerged for deliberation latency. The creation times this requires come from a later re-fetch that matches 39{,}920 of the 40{,}339 human pull requests in the 282 repositories matched on the records as fetched (99.0\%), and the others are dropped. Review and comment counts carry no timestamps, so for review rounds, comment volume, and changes requested, both arms are restricted to pull requests resolved by the cutoff.

Every latency comparison is thus conditional on resolution by the cutoff, and because we fit no survival model, it does not describe the longest waits. A common cutoff also does not equalize creation dates. Pooled over the resolved pull requests of the matched latency sample, the median agent pull request was created on 11 June 2025 and the median human pull request on 24 June 2025, so the median human pull request had 36 days in which to be resolved, against 50 for the median agent pull request. Within repositories, the median creation dates of the two arms differ by less than a day in the median repository, but by more than three weeks in either direction in half of the repositories.

Both arms use the same definitions. Comment volume counts issue comments only, because inline review comments were not fetched for human pull requests, and engagement means any review or comment in either arm. Conflict incidence has no human counterpart, since the replay cannot be reconstructed for historical human pull requests. The \emph{matched set} for a measure contains the repositories with at least ten pull requests in both arms after pull requests with a missing outcome or covariate are dropped. Matching is at the repository level, so the arms share repositories and repository covariates, but individual pull requests are not paired. Of the 404 repositories, 33 could no longer be accessed, and 318 had at least ten human pull requests (314 with at least ten resolved ones) before the removal of the remaining bot and agent accounts. Requiring ten agent pull requests of the relevant kind after the cap, with complete covariates, yields the matched sets of Table~\ref{tab:samples}. Each arm is fitted separately with the covariates available to both (pull-request size, the $M_2$ repository covariates, codebase size, and age in $M_3$, then team size and continuous integration in $M_4$), and the repository covariates, including activity, take the same value in both arms.

Three checks address alternative explanations in which agent pull requests are shaped differently, merged automatically, or different before any review, so that an agent-human difference would arise without the repository playing any part. A task-shape model adds \emph{diff dispersion}, the number of changed files and of distinct top-level directories a pull request touches. A fast-merge model adds an indicator for pull requests merged within two minutes without any review or comment. A placebo comparison repeats the analysis on pull-request size, which the author sets before the change meets the repository's review process. To examine the variance ratio itself, we refit the aligned comparison of resolution latency without each repository in turn, on three alternative outcome scales, with two treatments of creation date, and with two changes to the size of the arms (Section~\ref{sec:rq2} and Table~S4).

\subsection{Proxies for agent-specific explanations}
\label{sec:rivals}
Separate models for the two arms do not remove shared confounding entirely, and they leave open attributes that apply to agent pull requests alone. Public records allow proxies for two of them. A repository may treat agent pull requests by a rule of its own, and the developer who launches an agent may be a maintainer who merges its output directly. We measure a proxy for each, and a proxy for base-branch activity, compute each identically for both arms on the repositories of the matched comparisons, and add them to $M_4$ to form $M_5$.

A repository's \emph{stated policy toward agents} is proxied by two indicators dated from the file history, so that each pull request is coded with the files in force at its creation. The first records a contributing guide, pull-request template, or \texttt{CODEOWNERS} file that states a rule for AI agents or generated code, coded on the dimensions that catalogs of such policies use (permission, disclosure, and required human involvement)~\cite{hora2026landscape,hora2026disclosure,manita2026regulating}. The second records an agent context file such as \texttt{AGENTS.md} or \texttt{CLAUDE.md}~\cite{chatlatanagulchai2025readmes}, which shows that the repository has configured an agent without stating a rule about agent contributions. \emph{Launcher standing} is the relation to the repository, as GitHub's \texttt{author\_association} field reports it, of the account responsible for a pull request. For a human pull request, that account is the author. For an agent pull request, it is the launcher, whose account opens the pull request for Codex, Cursor, and Claude Code, and who is the assignee, issue author, or co-author that dispatched Copilot or Devin where the records identify one. Otherwise, the pull request is unattributed. \emph{Base-branch tempo} is the logarithm of the number of commits that landed on the target branch while the pull request was open. Because that count grows with the time the pull request stays open, we also count the commits in a fixed seven-day window after creation. Tempo is undefined when the target branch has since been deleted or the creation time is unknown, so $M_5$ is fitted on the pull requests with a defined tempo and compared with $M_4$ refitted on the same pull requests. A variant without tempo is fitted on the whole matched set.

\subsection{Reproducibility and data availability}
\label{sec:availability}
All analyses use publicly released datasets and read-only GitHub API queries.
The replication package contains the pipeline with fixed seeds, dataset hashes, the released covariates as deciles keyed to hashed repository identifiers, an audit script that checks the reported numbers against the released result tables, the scripts that draw the figures, and the supplementary material cited in this paper%
. The supplementary material also accompanies this preprint. The package for the first version of this preprint is archived on Zenodo (doi:10.5281/zenodo.20759752), and the package for the present version is prepared as a new version of that record.

\section{Results}
\label{sec:results}

\subsection{RQ1: How much do repositories differ for agent pull requests?}
\label{sec:rq1}

\begin{figure}[t]
\centering
\includegraphics[width=\columnwidth]{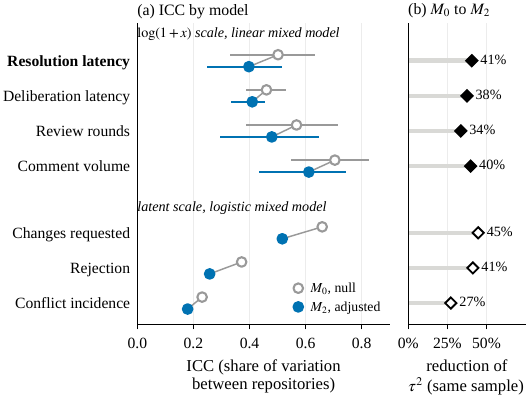}
\caption{RQ1, agent pull requests in the primary sample (repositories with $\ge$10 agent PRs; resolution latency on 26{,}694 PRs in 403 repositories). (a) ICC under the null model $M_0$ (open) and under $M_2$, which adds PR, agent, and repository covariates (filled). Bars are 95\% cluster-bootstrap intervals for the $\log(1+x)$ measures, with $B=1{,}000$ for resolution latency, 500 for deliberation latency, and 300 for review rounds and comment volume. (b) Reduction of the between-repository variance $\tau^2$ from $M_0$ to $M_2$ on the same sample. Latent-scale values come from logistic models and are not comparable with the $\log(1+x)$ values. $M_3$ (size and age) changes each ICC by at most 0.02 (Table~S2).}
\label{fig:rq1}
\end{figure}

\textbf{About two-fifths of the variance of the resolution latency of agent pull requests lies between repositories after the observable attributes of the contribution and the repository are accounted for.} In the 403 repositories of the primary sample with resolved pull requests (26{,}694 pull requests), half of the variance of resolution latency lies between repositories under the null model (ICC 0.50, 95\% CI [0.33, 0.63]). The $M_2$ covariates, which describe the pull request (size, task type, creation date, and engagement), its agent, and the repository's popularity, language, and activity, lower the ICC to 0.40 [0.25, 0.51]. Computed on the same pull requests, they remove 41\% of the between-repository variance $\tau^2$ and leave 59\% of it (Figure~\ref{fig:rq1}b). That is, they account for 41\% of the differences between repositories. On the 392 repositories for which the API returned both codebase size and age, the null-model ICC is again 0.50, and adding the two attributes ($M_3$) removes a further 0.7\% to 0.8\% of $\tau^2$, depending on whether they enter as raw values or as the released deciles (Table~S2), and leaves the ICC at 0.39 [0.24, 0.52].

The pattern is not specific to waiting time (Figure~\ref{fig:rq1}). Among the $\log(1+x)$ measures, the $M_2$ ICC is 0.41 for deliberation latency, 0.48 for review rounds, and 0.61 for comment volume. Among the latent-scale measures, it is 0.52 for changes requested, 0.26 for rejection, and 0.18 for conflict incidence. For every measure, the covariates remove between 27\% and 45\% of $\tau^2$, and a between-repository share remains. Codebase size and age remove less than 3\% of the remaining $\tau^2$ for five of the seven measures and at most 11.5\%, for conflict incidence (Table~S2). Among the latent-scale measures, the share is highest for whether a pull request draws any review or comment at all, with an ICC for engagement of 0.85 under $M_0$ and 0.78 under $M_3$. Deliberation latency, which is defined only for engaged pull requests, therefore depends on a selection whose rate differs between repositories. Resolution latency, by contrast, is defined for all resolved pull requests and gives a similar ICC under $M_3$ (0.39 against 0.41).

The share could depend on the sample threshold or on a few unusual repositories, and it persists across both checks. Across activity cutoffs of 1, 5, 10, and 20 agent pull requests per repository, the $M_2$ ICC of resolution latency ranges from 0.40 to 0.52 and is highest when repositories with a single agent pull request are included (Figure~S1). A leave-one-repository-out jackknife of the $M_3$ ICC gives values between 0.31 and 0.40 around the full estimate of 0.39 (Figure~S2). Removing one repository lowers the estimate to 0.31, and removing another lowers it to 0.36. None of the other 390 removals moves it by 0.01 or more. The between-repository share therefore persists without any single repository, although one repository carries a visible part of it.

Repositories that received pull requests from several agents, where agent contributions could interfere with one another, do not show a larger repository share. With activity in the model, the $M_2$ ICC of resolution latency is 0.21 [0.12, 0.31] in the 107 repositories with two or more agents and 0.47 [0.26, 0.60] in the 285 with one, with the same ordering for deliberation latency (0.34 against 0.43). Across the curated subset, the conflict rate is similar in repositories with several agents and in repositories with one (30.8\% against 31.2\%), and in AgenticFlict it ranges from 15.4\% for Copilot to 32.3\% for Codex across the five agents. The number of agents is a coarse measure of concurrent agent activity, because one agent product can run many instances through one account, so these comparisons leave open whether agent contributions interfere with one another.

\subsection{RQ2: Do repositories differ more for agent than for human pull requests?}
\label{sec:rq2}

Repositories also differ for human pull requests~\cite{yu2015,zhang2022prlatency}, so whether the RQ1 share is particular to agents can only be judged against human pull requests in the same repositories, observed up to the same date.

\begin{figure}[t]
\centering
\includegraphics[width=\columnwidth]{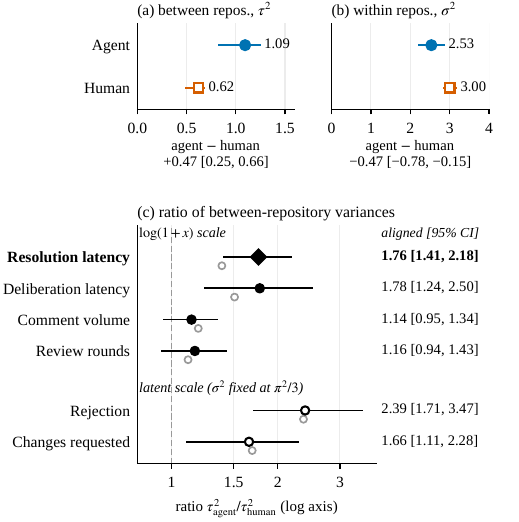}
\caption{RQ2, matched repositories, common covariates ($M_3$), aligned observation cutoff. (a, b) Resolution latency on the $\log(1+\text{hours})$ scale, 280 repositories, 10{,}623 agent and 36{,}102 human PRs: between- and within-repository variance of each arm with 95\% percentile intervals, and the agent-minus-human difference, from a cluster bootstrap over repositories ($B=1{,}000$). (c) Ratio of between-repository variances with 95\% intervals (black) and the estimate on the records as fetched (grey); a ratio above 1 means that repositories differ more for agent PRs. Filled markers show the latencies on $\log(1+\text{hours})$ and the counts on $\log(1+\text{count})$, and open markers show rejection and changes requested on the latent scale, with $\sigma^2$ fixed. Ratios on different scales are not directly comparable.}
\label{fig:rq2}
\end{figure}

\textbf{In the same repositories, the between-repository variance of resolution latency is larger for agent pull requests than for human pull requests, and the within-repository variance is smaller.} In the 280 matched repositories (10{,}623 agent and 36{,}102 human pull requests resolved by the cutoff), under the common covariates of $M_3$, the between-repository variance is 1.09 for agent pull requests and 0.62 for human pull requests. The difference is 0.47 (95\% CI [0.25, 0.66]), and the ratio is 1.76 [1.41, 2.18] (Figure~\ref{fig:rq2}a). On the $\log(1+\text{hours})$ scale, the standard deviation of the repository offsets is 1.05 for the agent arm and 0.79 for the human arm. For waits of several hours, where $\log(1+\text{hours})$ is close to $\log(\text{hours})$, a repository one standard deviation slower than the average therefore multiplies the typical resolution latency of an agent pull request by about 2.8 and that of a human pull request by about 2.2. The within-repository variance is 2.53 for agent pull requests and 3.00 for human pull requests, a difference of $-0.47$ [$-0.78$, $-0.15$] (Figure~\ref{fig:rq2}b), so agent pull requests in one repository are less dispersed in resolution latency than human ones there. These estimates concern how much repositories differ, among resolved pull requests, and do not compare average waiting times.

Both components push the ICC in the same direction. The ICC of the agent arm is 0.30 against 0.17 for the human arm, an \emph{ICC gap} (agent minus human) of $+0.13$ [0.08, 0.18] (Table~\ref{tab:rq2}). Because an ICC can rise through a larger between-repository variance or through a smaller within-repository variance (Section~\ref{sec:record}), we allocate the gap to the two components by giving both arms the same within-repository variance. The allocation depends on which arm's value is used. With the human value, the agent ICC would be 0.27, so about three-quarters of the gap would remain and be due to the between-repository variance. With the agent value, the human ICC would be 0.20 ($0.621/(0.621+2.534)$), and about four-fifths of the gap would remain. Either way, most of the gap reflects the larger between-repository variance of the agent arm. The smaller within-repository variance may reflect fewer launching accounts or agent products within each repository, which this comparison does not test.

Observing both arms up to the same date matters. On the records as fetched, whose human outcomes run to June 2026, the ratio is 1.39 [1.07, 1.76] and the within-repository difference is $-1.50$ [$-1.83$, $-1.14$] on 282 repositories (Table~S3). The cutoff reclassifies 3{,}163 of the 39{,}271 resolved human pull requests with a known creation time (8.1\%) as still open and lowers both human components, the within-repository variance from 4.05 to 3.00 and the between-repository variance from 0.78 to 0.62, whereas dropping the human pull requests without a creation time leaves the ratio at 1.40. Requiring at least 30 days of follow-up for every pull request gives a ratio of 1.98 [1.42, 2.67] on 173 repositories, and 1.68 [0.90, 3.11] for deliberation latency on 76 repositories, an interval that includes one. Alignment therefore increases the between-repository ratio and reduces the within-repository difference by about two-thirds, although the interval of that difference still excludes zero.

\begin{table}[t]
\centering
\caption{RQ2 ICCs on the matched repositories ($M_3$, common covariates), aligned observation cutoff. Repositories: 280 (10{,}623 agent and 36{,}102 human PRs) for every measure except deliberation latency (138; 4{,}341 and 15{,}065). For review rounds, comment volume, and changes requested, both arms are restricted to PRs resolved by the cutoff. Gap = agent minus human with a 95\% cluster-bootstrap CI ($B=1{,}000$); gaps use unrounded ICCs. $\dagger$ latent scale.}
\label{tab:rq2}
\footnotesize
\rowcolors{2}{gray!12}{white}
\begin{tabular}{@{}p{0.36\columnwidth}ccc@{}}
\toprule
Measure & Agent & Human & Gap [95\% CI] \\
\midrule
Resolution latency & 0.30 & 0.17 & $+0.13$ [0.08, 0.18] \\
Deliberation latency & 0.43 & 0.20 & $+0.22$ [0.14, 0.29] \\
Review rounds & 0.39 & 0.31 & $+0.08$ [0.02, 0.13] \\
Comment volume & 0.57 & 0.48 & $+0.09$ [0.02, 0.15] \\
Changes requested$^\dagger$ & 0.55 & 0.43 & $+0.13$ [0.03, 0.20] \\
Rejection$^\dagger$ & 0.27 & 0.13 & $+0.14$ [0.08, 0.19] \\
\bottomrule
\end{tabular}
\end{table}

\textbf{The larger between-repository variance extends to whether and when a pull request is integrated, and to whether changes are requested, but not detectably to the amount of review activity.} The interval of the aligned variance ratio excludes one for resolution latency, deliberation latency (1.78 [1.24, 2.50]), rejection (2.39 [1.71, 3.47]), and changes requested (1.66 [1.11, 2.28]), and includes one for review rounds (1.16 [0.94, 1.43]) and comment volume (1.14 [0.95, 1.34]) (Figure~\ref{fig:rq2}c). On the records as fetched, the interval for comment volume excluded one (1.19 [1.01, 1.38]). Engagement, which rests on the same untimed counts, was not re-estimated at the cutoff, and on the records as fetched, its latent-scale ICC is 0.77 for agent and 0.67 for human pull requests. When the agent arm is given the human within-repository variance, between 0.38 and 0.73 of each aligned ICC gap among the $\log(1+x)$ measures is attributed to the between-repository variance (Table~S3). For the two latent-scale measures, the within-repository variance is fixed at $\pi^2/3$ in both arms, so the whole ICC gap is between-repository variance by construction. The agent ICC of 0.30 comes from a different sample and model than the RQ1 value of 0.39, since it uses the 200 most recent agent pull requests of each matched repository and the common covariates, and the two values are not interchangeable (Section~\ref{sec:threats}). Adding team size and continuous integration ($M_4$) changes each aligned ICC gap of the $\log(1+x)$ measures by at most 0.02, and for resolution latency the ratio becomes 2.00 [1.56, 2.49]. The interval of every aligned ICC gap in Table~\ref{tab:rq2} excludes zero.

A variance ratio can depend on how the outcome is scaled and on which pull requests enter each arm, so we refit it under alternative specifications (Table~S4). \textbf{The ratio for resolution latency exceeds one, with an interval that excludes one, in every specification on a log or rank scale, while its size depends on the scale and on the calendar composition of the arms.} On those scales, the point estimates range from 1.47 to 1.84 (1.56 to 1.93 across 50 random equal-size subsamples). The lowest values come from a rank-based inverse-normal transform, which replaces each wait by a normal score of its rank (1.47 [1.19, 1.83]), and from restricting each repository's two arms to the overlap of their creation dates (1.49 [1.14, 2.01] on 183 repositories), and the highest from adding creation date to $M_3$ (1.84 [1.46, 2.28]). Leaving out one repository at a time gives ratios between 1.71 and 1.80, and subsampling each repository's larger arm to the size of the smaller one gives 1.80 [1.42, 2.31]. On raw hours, whose variance is dominated by a few very long waits, the ratio is 1.89 with an interval from 0.83 to 3.75. The within-repository difference is less stable, since its interval includes zero both under the creation-date overlap ([$-0.78$, 0.05]) and with equal arm sizes ([$-0.54$, 0.06]). The direction of the between-repository difference is thus stable on log and rank scales, whereas the smaller within-repository variance of agent pull requests is tentative.

The three checks of Section~\ref{sec:baseline} were run on the records as fetched. Agent and human pull requests change a similar number of files (median 3 in both arms, standardized mean difference $-0.004$), whereas agent pull requests touch slightly more top-level directories (standardized mean difference 0.21) and change more lines (0.27). Adding diff dispersion moves each ICC gap by at most 0.006, and every interval still excludes zero. In the resolution-latency sample, 9.6\% of agent and 2.9\% of human pull requests were merged within two minutes with no recorded review or comment (a \emph{fast merge without engagement}), and 14.8\% and 6.8\% within two minutes with or without engagement. Adding the fast-merge indicator leaves the ICC gap at 0.14 in the records as fetched. Because the indicator is partly defined by resolution latency, this adjustment does not identify the contribution of fast merges to the gap. On pull-request size, the placebo outcome, the ICC gap is $+0.10$ [0.05, 0.15] for changed lines and $+0.09$ [0.03, 0.13] for changed files. Part of the difference between the arms may therefore exist before contributions meet the review process. The latency gaps lie above the placebo gaps, but their intervals overlap and the difference has not been tested directly.

\subsection{RQ3: Do measured explanations account for the difference?}
\label{sec:rq3}

\textbf{The proxies for stated policy and launcher standing remove little of the agent arm's between-repository variance, and for resolution latency the agent-human difference remains when the proxies are added.}

The descriptives use the records as fetched for the 317 repositories that enter any matched comparison (11{,}867 agent and 44{,}878 human pull requests, resolved or not). The policy, context-file, and launcher codes, which require a creation time and a readable file history, are defined for 11{,}854 agent and 44{,}457 human pull requests in 316 of these repositories. Repositories configure agents more often than they state rules for them. An agent context file was in force in 169 repositories (53.5\%), for 44.0\% of agent and 42.4\% of human pull requests. In 7 of the 316 repositories (2.2\%), a contribution policy toward agents was in force when at least one agent pull request was created, and it applied to 2.2\% of agent and 2.7\% of human pull requests.

Agent pull requests are also launched more often by owners or members of the repository. A launcher is identified for 80\% of agent pull requests. The unattributed rest are almost all Devin's, since 81\% of Devin's pull requests lack an identifiable launcher, against 0.1\% or less of any other agent's. Owners or members launched 43.8\% of agent pull requests, collaborators or contributors 32.1\%, and outsiders without a formal association 3.7\%, against 27.8\%, 67.4\%, and 4.8\% of human pull requests. Tempo is undefined for 10.9\% of the 11{,}867 agent pull requests and for 5.0\% of the 44{,}457 human pull requests with a known creation time. The affected agent pull requests are not a random subset, since 84\% of them were merged, against 60\% of the others. The proxy models then keep, within each matched set, the pull requests with defined proxies, which leaves 260 repositories for resolution latency with tempo and 280 without it (Table~\ref{tab:samples}).

Table~\ref{tab:rq3} reports what each proxy removes. The policy and context-file indicators change the agent arm's $\tau^2$ by less than 2.5\% in either direction for every measure. Launcher standing removes 10\% of $\tau^2$ for rejection (11\% without tempo) and 5\% for comment volume, and changes it by less than 3\% elsewhere. As measured, launcher standing thus accounts for at most about a tenth of the differences between repositories. Without the tempo proxy, at the aligned cutoff, the variance ratio for resolution latency under $M_5$ is 2.13 [1.66, 2.69] on 280 repositories, against 2.00 [1.56, 2.49] under $M_4$ on the same repositories. We treat this model as the main RQ3 estimate for resolution latency, because the open-window tempo depends on the outcome.

\begin{table*}[t]
\centering
\caption{RQ3. Share of the agent arm's $M_4$ between-repository variance removed by each proxy alone and by all four ($M_5$), in \% (negative = increased), on the tempo-complete matched sample (resolution latency and rejection: 260 repositories and 9{,}191 agent PRs; deliberation latency: 129; other measures: 294), which the observation cutoff of the human arm does not change. Variance ratio with 95\% cluster-bootstrap CI ($B=1{,}000$) under $M_5$ with the open-window tempo on that sample, and without tempo on the whole matched set (resolution latency 280 repositories, deliberation latency 138, rejection 281, other measures 316). Latency ratios at the aligned observation cutoff; $\ddagger$ on the records as fetched; $\dagger$ latent scale.}
\label{tab:rq3}
\footnotesize
\rowcolors{2}{gray!12}{white}
\begin{tabular}{@{}lrrrrrcc@{}}
\toprule
Measure & Policy & Context & Launcher & Tempo & All & Ratio, $M_5$ & Ratio, $M_5$ without tempo \\
\midrule
Resolution latency & $-0.4$ & $-0.7$ & $-2.4$ & 20.4 & 18.4 & 2.18 [1.44, 3.09] & 2.13 [1.66, 2.69] \\
Deliberation latency & $-0.5$ & $-0.2$ & $-0.5$ & 28.5 & 27.3 & 1.61 [1.09, 2.20] & 2.47 [1.58, 3.62] \\
Review rounds$^\ddagger$ & 0.0 & 2.3 & $-1.0$ & 2.5 & 2.9 & 1.25 [1.00, 1.57] & 1.29 [1.04, 1.58] \\
Comment volume$^\ddagger$ & $-0.3$ & 0.0 & 5.1 & 2.3 & 7.0 & 1.13 [0.93, 1.31] & 1.12 [0.94, 1.32] \\
Changes requested$^{\dagger\ddagger}$ & 0.0 & $-0.4$ & 1.5 & 3.5 & 4.3 & 1.71 [1.14, 2.48] & 1.78 [1.22, 2.44] \\
Rejection$^{\dagger\ddagger}$ & 0.0 & 0.1 & 10.0 & $-0.5$ & 7.9 & 2.15 [1.33, 3.26] & 2.01 [1.31, 3.00] \\
\bottomrule
\end{tabular}
\end{table*}

Tempo over the open window removes 20.4\% of $\tau^2$ for resolution latency and 28.5\% for deliberation latency, but the count grows with the wait it is meant to explain. Its coefficient is $+1.11$ log-hours per log-commit in the agent arm and $+1.23$ in the human arm at the cutoff, and on the 260 tempo-complete repositories the agent arm's within-repository variance falls from 2.62 to 0.93 when tempo enters, so the open-window count partly restates the outcome. With it, the ratio for resolution latency on the 260 tempo-complete repositories is 2.10 [1.57, 2.65] under $M_4$ and 2.18 [1.44, 3.09] under $M_5$ (Table~\ref{tab:rq3}). With the seven-day count in place of the open-window count, the four proxies together remove 1.5\% of the agent arm's $\tau^2$ for resolution latency. The seven-day coefficient is negative in both arms ($-0.23$ for agent pull requests, which the cutoff does not affect, and $-0.27$ for human pull requests on the records as fetched), so pull requests opened while the base branch moves fast were resolved sooner. No interval was computed for these coefficients, and because the window follows creation, they describe an association, not a prospective effect. For the measures estimated on the records as fetched, the $M_5$ interval excludes one for changes requested and rejection, has a lower bound of 1.00 for review rounds (1.04 without tempo), and includes one for comment volume.

A stated policy was in force in 7 of 316 repositories, too few for its indicator to remove much variance, so the result does not show that policies are ineffective. A policy file records a stated rule whose enforcement the data do not show, launcher attribution is missing for one agent pull request in five, and the data lack other agent-specific attributes. The between-repository variance that remains after the proxies therefore has no identified source.

\section{Discussion}
\label{sec:discussion}

\subsection{Implications for overseeing coding agents}
\label{sec:implications}
The OpenAI incident showed that agents pursuing their own tasks can coordinate through the infrastructure they share~\cite{metr2026incident}. Our results show that, in everyday software development, the outcomes of agent pull requests also vary with the receiving repository, more than those of human pull requests (RQ1 and RQ2). Three implications follow.

First, evidence about agents needs the repository as a unit. Pooled summaries such as acceptance rates or merge times per agent~\cite{aiteammates2025} also reflect differences between repositories. Because 285 of the 392 repositories of RQ1 received pull requests from a single agent, a comparison of agents that pools pull requests across repositories cannot separate an agent from the repositories in which it works. Pull requests from one repository are also correlated (for resolution latency, an ICC of 0.39), so analyses that treat them as independent understate the uncertainty of comparisons between agents~\cite{raudenbush2002,snijders2012}. Multilevel models address this second point, whereas separating agents from repositories requires comparing agents within the same repositories, which the 107 of the 392 repositories with several agents allow in principle.

Second, oversight needs signals at the level of the repository. After its incident, OpenAI announced more isolated sandboxes, more monitoring of the agents' reasoning, and wider use of graders that assess how a task was completed~\cite{openai2026incident}. A repository has counterparts in its review rules and merge settings, and our data show what they would need to watch. On the records as fetched, agent pull requests were merged within two minutes without any recorded review or comment about three times as often as human ones (9.6\% against 2.9\%), and whether a pull request draws any review or comment differs between repositories (latent-scale ICC 0.77 for agent pull requests). A project can track such quantities from records it already keeps and compare its agent pull requests with its own human ones.

Third, few repositories state rules for agent contributions. Agent context files were in force in 53.5\% of the repositories and a stated policy toward agent contributions in 7 of 316 (Section~\ref{sec:rq3}), so nearly all agent pull requests were created where no stated policy for agents was in force.

\subsection{Explanations consistent with the results}
Several explanations fit RQ1 to RQ3, and each needs data that public records lack. First, a repository may apply review rules, merge permissions, or automation to agent pull requests without stating them in a policy file. Repositories that differ in such unstated rules would produce a larger between-repository variance for agent pull requests, while a policy indicator would remove little of it, as RQ3 finds. Testing this requires review assignments, branch-protection settings, and merge permissions recorded over time. Second, launchers may run agents in repositories they control and merge the output themselves. Owners or members launch agent pull requests more often than they author human ones (Section~\ref{sec:rq3}), yet launcher standing removes little of $\tau^2$, and one agent pull request in five has no identified launcher, so a test needs the launching and merging account of every agent pull request. Third, most repositories received pull requests from a single agent, and the common covariates of RQ2 omit agent identity, so part of the agent arm's between-repository variance, and its smaller within-repository variance, may reflect which agent a repository uses. A test requires several agents working in the same repositories. Fourth, interactions among contributors, which a collective account of AI-native repositories proposes as a source of repository-level outcomes~\cite{emergencetheory}, are consistent with the remaining variance. In public repositories, however, those with several agents show a smaller repository share (Section~\ref{sec:rq1}), and base-branch tempo is only a mechanical proxy for interaction. An intervention on merge scheduling or review routing would test this explanation more directly than further observational records.

\subsection{Local evaluation and project controls as hypotheses}
\label{sec:local}
Because resolution latency varies more between repositories for agent pull requests, the latency they meet in one repository is a less certain guide to the latency they meet in another. We therefore hypothesize that resolution latency measured during a fixed trial in the target repository, against the repository's own human pull requests, predicts later integration outcomes, such as reverts, rework, and abandoned pull requests, better than the agent's benchmark score or prior track record. A prospective study would compare a model of those outcomes built on benchmark standing and track record with one that adds the typical value and spread of the agent's resolution latency in the first weeks of use. Merge queues for frequently changed modules, limits on the rate of agent submissions, and rules that route agent pull requests through review act on the quantities of Section~\ref{sec:implications}. This study did not evaluate them, and each needs an intervention study before it is recommended as practice.

\section{Threats to Validity}
\label{sec:threats}

\textbf{Construct validity.} Resolution latency counts automated resolutions together with human ones, so it measures how long a contribution waited, not how long anyone deliberated over it, the size of a queue, or the effort of a reviewer. Pull requests still open at the cutoff are excluded in both arms, which removes the longest waits and, if repositories differ in how many they leave open, also changes the between-repository variance. The $\log(1+\text{hours})$ transform compresses waits under an hour, where many agent merges lie, and the resolution-latency ICC depends on it, at 0.39 on $\log(1+\text{hours})$, 0.36 on raw hours, and 0.31 after a rank-based inverse-normal transform under $M_2$ on 392 repositories. The size of the aligned variance ratio depends on it too, but not its direction on log and rank scales (Section~\ref{sec:rq2}). Part of the agent arm's smaller within-repository variance may reflect the mass of waits near zero, although deliberation latency, which excludes the fastest merges, shows the same ordering. Human authorship is inferred by exclusion, so undetected agent-authored or automated pull requests may remain in the human arm. Because their effect depends on how they are distributed across repositories, the direction of the resulting bias is unknown.

\textbf{Internal validity.} The design is observational, so the between-repository variance that remains after the covariates establishes no cause. Holding the repository fixed makes the two arms more comparable, but the arms are fitted as separate models, and pull requests are not paired. Shared confounding and selection can therefore persist, including from the agents a repository uses, since agent identity is not a common covariate. The common cutoff aligns the end of observation without aligning creation dates, and adjusting for creation date or restricting to overlapping creation dates changes the size of the ratio but not its direction (Section~\ref{sec:rq2}). Review and comment counts of human pull requests carry no timestamps, so comments posted after the cutoff remain in the human counts until a re-fetch of those timestamps puts the count measures on the same cutoff. Keeping the 200 most recent pull requests per repository gives both arms many recent pull requests with a short time to be resolved, and requiring 30 days of follow-up raises the resolution-latency ratio (Section~\ref{sec:rq2}). Governance norms, launcher practices, and application domain remain uncontrolled, and codebase size was measured in June 2026, after the outcomes.

\textbf{Conclusion validity.} The arms differ in size, with 3.4 human pull requests for every agent pull request in the aligned resolution-latency sample, and the agent ICC depends on the cap, since, on the records as fetched, it is 0.46 without the cap and 0.30 with the most recent or a random 200 pull requests per repository. The latent-scale ICCs depend on the logistic $\pi^2/3$ convention~\cite{goldstein2002,nakagawa2017}, and linear-probability fits agree with them in the direction of every agent-human gap. Bootstrap replicates of the binary models use a 200-point integration grid, which agrees with the 400-point grid of the point estimates to within 0.001 on $\tau^2$.

\textbf{External validity.} The study covers public repositories with more than 100 stars, five agents, and the first seven months of 2025~\cite{aidev2026,agenticflict2026}, and it trades control over the setting for observation of real contributions in their own workflows~\cite{stol2018}. Mining GitHub carries known hazards, among them unrepresentative repository samples and merges that happen outside the platform~\cite{kalliamvakou2016perils,nagappan2013diversity}, so the results may not extend to closed industrial repositories, to smaller projects, or to later agent versions.

\section{Conclusion}
\label{sec:conclusion}

Agents that share infrastructure can act in ways that evaluations of single agents do not reveal, as the OpenAI incident showed at its extreme~\cite{metr2026incident}. For coding agents in public repositories, about two-fifths of the variation in the resolution latency of pull requests lies between repositories, and in the same repositories the between-repository variance is larger for agent than for human pull requests (variance ratio 1.76, 95\% CI 1.41 to 2.18), with the same direction for rejection. Its direction holds on log and rank scales, and the measured proxies do not account for it. The repository therefore belongs in the model whenever agent outcomes are compared across repositories, and it should be part of the oversight of coding agents. Whether local measurement predicts an agent's later outcomes better than its benchmark score remains to be tested.

\bibliographystyle{IEEEtran}
\bibliography{references}

\end{document}

\begin{center}
{\large\bfseries Supplementary Material}\\[3pt]
{\itshape For Coding Agents, the Repository Matters}\\[2pt]
{\small Daniel Russo, Department of Computer Science, Aalborg University, Copenhagen, Denmark}

\end{center}

This document collects the descriptives, robustness tables, and figures that the paper cites as Tables S1 to S4 and Figures S1 and S2. Every number traces to a table in \texttt{results/tables/} of the replication package, named under each exhibit, and every figure is drawn by a script in \texttt{src/} from those tables.

\begin{table}[!htbp]
\centering
\caption{Setting descriptives of the agent pull requests (PRs). Figures are computed on the curated AIDev subset (2{,}807 repositories with more than 100 stars, 33{,}596 PRs) unless a row notes the full release (116{,}211 repositories). Source: \texttt{rq1\_summary.json}, \texttt{rq1\_multiagent\_share\_by\_stratum.csv}.}
\label{tab:setting}
\small
\begin{tabular}{p{0.62\textwidth}r}
\toprule
Measure & Value \\
\midrule
PRs with a review or comment from an account other than the author & 39.7\% \\
Repositories with at least one such PR & 66.5\% \\
PRs with at least one formal review (review events per reviewed PR) & 24.2\% (3.5) \\
PRs with at least one comment & 38.6\% \\
Reviews that requested changes & 1{,}604 \\
Repositories with PRs from $\ge 2$ of the five agents (curated / full release) & 8.3\% / 1.7\% \\
\quad among curated repositories with $\ge 20$ agent PRs & 31.2\% \\
Repositories with several agents whose PRs from different agents fall within 7 days & 49.8\% \\
PRs covered by the AgenticFlict replay & 28.5\% \\
Conflict rate, repositories with several agents against one agent & 30.8\% against 31.2\% \\
Directly observable agent-to-agent actions (lower bound) & 2 \\
\bottomrule
\end{tabular}
\end{table}

The count of direct agent-to-agent actions is a lower bound because Codex, Cursor, and Claude Code act through the operator's own account, so an action by one of them on another agent's PR is recorded as an action by a person.

\begin{table}[!htbp]
\centering
\caption{Codebase size and repository age (RQ1). ICC under $M_2$ and under $M_3=M_2+{}$size${}+{}$age on the repositories of the primary sample whose GitHub-API attributes could be fetched (392 for resolution latency and rejection, 393 for review rounds, comment volume, and changes requested, 336 for deliberation latency, and 330 for conflict incidence), with size and age entered as deciles, the released form. The $M_2$ values differ from Figure~1 of the paper because the sample does. The last column is the share of the $M_2$ between-repository variance that size and age remove. $\dagger$ latent scale. Source: \texttt{rq2\_variance\_partition\_m3.csv} (the raw-log form, with a maximum share of 13.9\%, is in the same file).}
\label{tab:confound}
\small
\begin{tabular}{lccc}
\toprule
Measure & ICC $M_2$ & ICC $M_3$ & Share removed by size and age \\
\midrule
Resolution latency & 0.39 & 0.39 & 0.8\% \\
Deliberation latency & 0.42 & 0.41 & 2.6\% \\
Review rounds & 0.48 & 0.48 & 0.7\% \\
Comment volume & 0.61 & 0.61 & 1.1\% \\
Changes requested$^\dagger$ & 0.52 & 0.52 & 0.7\% \\
Rejection$^\dagger$ & 0.25 & 0.23 & 9.2\% \\
Conflict incidence$^\dagger$ & 0.18 & 0.16 & 11.5\% \\
\bottomrule
\end{tabular}
\end{table}

\begin{table}[!htbp]
\centering
\caption{Variance components of the matched comparison (RQ2), common covariates of $M_3$ and $M_4$. $\tau^2$ is the between-repository and $\sigma^2$ the within-repository variance, on the $\log(1+x)$ scale or, for latent-scale measures ($\dagger$), on the latent scale of a logistic model, where $\sigma^2$ is fixed at $\pi^2/3\approx3.29$. Ratio $=\tau^2_{\text{agent}}/\tau^2_{\text{human}}$ with a 95\% cluster-bootstrap interval ($B=1{,}000$). The last column is the share of the ICC gap that the between-repository variance accounts for when the agent arm is given the human $\sigma^2$. Specifications: \emph{cut.}, the aligned observation cutoff, the latest closing event in the agent records (30 July 2025, 23:20 UTC), applied to the human arm, so that a human PR resolved later counts as still open; \emph{cut., res.}, as cut., with both arms restricted to PRs resolved by the cutoff, because the review and comment counts carry no timestamps; \emph{no cut.}, human PRs with a known creation time, without the cutoff; \emph{30 d}, as cut., with both arms restricted to PRs created at least 30 days before the cutoff; \emph{fetch.}, the records as fetched, whose human outcomes run to June 2026. In every specification, the latency and rejection models use resolved PRs only, and no survival model is fitted. In the 282 repositories matched on the records as fetched, the cutoff reclassifies 3{,}163 of the 39{,}271 resolved human PRs with a known creation time as still open. Sources: \texttt{rq3\_horizon\_components.csv} (A, B; specifications \texttt{censored}, \texttt{censored\_resolved}, \texttt{aligned}, and \texttt{censored30}), \texttt{rq3\_variance\_components.csv} (C), and \texttt{rq3\_horizon\_counts.csv}; the rows are written by \texttt{src/tables\_supplement.py}.}
\label{tab:components}
\small
\setlength{\tabcolsep}{3.2pt}
\begin{tabular}{@{}llcrccccll@{}}
\toprule
& & & & \multicolumn{2}{c}{$\tau^2$} & \multicolumn{2}{c}{$\sigma^2$} & & Share of gap \\
\cmidrule(lr){5-6}\cmidrule(lr){7-8}
Measure & Spec. & Model & Repos & Agent & Human & Agent & Human & Ratio [95\% CI] & from $\tau^2$ \\
\midrule
\multicolumn{10}{@{}l}{\itshape A. Aligned observation cutoff (primary)} \\
Resolution latency & cut. & $M_3$ & 280 & 1.09 & 0.62 & 2.53 & 3.00 & 1.76 [1.41, 2.18] & 0.73 \\
Resolution latency & cut. & $M_4$ & 280 & 0.98 & 0.49 & 2.53 & 3.01 & 2.00 [1.56, 2.49] & 0.76 \\
Deliberation latency & cut. & $M_3$ & 138 & 1.13 & 0.64 & 1.53 & 2.51 & 1.78 [1.24, 2.50] & 0.49 \\
Deliberation latency & cut. & $M_4$ & 138 & 1.04 & 0.50 & 1.53 & 2.51 & 2.07 [1.41, 3.13] & 0.53 \\
Review rounds & cut., res. & $M_3$ & 280 & 0.19 & 0.16 & 0.30 & 0.36 & 1.16 [0.94, 1.43] & 0.42 \\
Review rounds & cut., res. & $M_4$ & 280 & 0.18 & 0.15 & 0.30 & 0.36 & 1.23 [0.99, 1.52] & 0.49 \\
Comment volume & cut., res. & $M_3$ & 280 & 0.23 & 0.20 & 0.18 & 0.22 & 1.14 [0.95, 1.34] & 0.38 \\
Comment volume & cut., res. & $M_4$ & 280 & 0.23 & 0.20 & 0.18 & 0.22 & 1.14 [0.94, 1.34] & 0.37 \\
Changes requested$^\dagger$ & cut., res. & $M_3$ & 280 & 4.08 & 2.46 & 3.29 & 3.29 & 1.66 [1.11, 2.28] & 1.00 \\
Rejection$^\dagger$ & cut. & $M_3$ & 280 & 1.22 & 0.51 & 3.29 & 3.29 & 2.39 [1.71, 3.47] & 1.00 \\
\midrule
\multicolumn{10}{@{}l}{\itshape B. Sensitivity of the alignment} \\
Resolution latency & no cut. & $M_3$ & 281 & 1.09 & 0.78 & 2.54 & 4.05 & 1.40 [1.09, 1.75] & 0.37 \\
Deliberation latency & no cut. & $M_3$ & 138 & 1.13 & 0.74 & 1.53 & 2.88 & 1.53 [1.06, 2.22] & 0.35 \\
Resolution latency & 30 d & $M_3$ & 173 & 1.22 & 0.61 & 2.59 & 3.43 & 1.98 [1.42, 2.67] & 0.66 \\
Deliberation latency & 30 d & $M_3$ & 76 & 0.93 & 0.56 & 1.45 & 2.83 & 1.68 [0.90, 3.11] & 0.37 \\
\midrule
\multicolumn{10}{@{}l}{\itshape C. Records as fetched} \\
Resolution latency & fetch. & $M_3$ & 282 & 1.09 & 0.78 & 2.54 & 4.04 & 1.39 [1.07, 1.76] & 0.36 \\
Resolution latency & fetch. & $M_4$ & 282 & 0.98 & 0.62 & 2.54 & 4.04 & 1.58 [1.22, 2.05] & 0.43 \\
Deliberation latency & fetch. & $M_3$ & 138 & 1.13 & 0.75 & 1.53 & 2.86 & 1.51 [1.05, 2.20] & 0.35 \\
Deliberation latency & fetch. & $M_4$ & 138 & 1.04 & 0.58 & 1.53 & 2.86 & 1.79 [1.20, 2.70] & 0.41 \\
Review rounds & fetch. & $M_3$ & 317 & 0.17 & 0.16 & 0.32 & 0.41 & 1.11 [0.92, 1.37] & 0.30 \\
Review rounds & fetch. & $M_4$ & 317 & 0.17 & 0.14 & 0.32 & 0.41 & 1.20 [0.99, 1.47] & 0.42 \\
Comment volume & fetch. & $M_3$ & 317 & 0.23 & 0.19 & 0.19 & 0.26 & 1.19 [1.01, 1.38] & 0.38 \\
Comment volume & fetch. & $M_4$ & 317 & 0.23 & 0.19 & 0.19 & 0.26 & 1.20 [1.02, 1.40] & 0.39 \\
Changes requested$^\dagger$ & fetch. & $M_3$ & 317 & 3.88 & 2.29 & 3.29 & 3.29 & 1.69 [1.18, 2.33] & 1.00 \\
Changes requested$^\dagger$ & fetch. & $M_4$ & 317 & 3.82 & 2.14 & 3.29 & 3.29 & 1.79 [1.22, 2.40] & 1.00 \\
Rejection$^\dagger$ & fetch. & $M_3$ & 282 & 1.21 & 0.51 & 3.29 & 3.29 & 2.36 [1.68, 3.32] & 1.00 \\
Rejection$^\dagger$ & fetch. & $M_4$ & 282 & 1.15 & 0.45 & 3.29 & 3.29 & 2.58 [1.77, 3.71] & 1.00 \\
\bottomrule
\end{tabular}
\end{table}

\begin{table}[!htbp]
\centering
\caption{Sensitivity of the aligned variance ratio of resolution latency (RQ2): common covariates of $M_3$, aligned observation cutoff, resolved PRs. Each row changes one element of the primary specification. $\tau^2$ and $\sigma^2$ as in Table~S3, on the scale of the row, so components in different rows are not comparable (raw-hours components in squared hours). Ratio $=\tau^2_{\text{agent}}/\tau^2_{\text{human}}$ with a 95\% cluster-bootstrap interval ($B=1{,}000$, seed 42); the last column is the 95\% interval of $\sigma^2_{\text{agent}}-\sigma^2_{\text{human}}$. Creation date enters linearly and quadratically, on a scale common to both arms. \emph{Overlapping creation dates} restricts each repository's two arms to the intersection of the creation-date ranges of their resolved PRs and then reapplies the rule of ten PRs per arm. The rank-based inverse-normal transform is computed on the pooled PRs of both arms. $^a$Minimum and maximum over the 280 removals; no removal moves the ratio by 0.10 or more. $^b$The larger arm of each repository is subsampled at random to the size of the smaller one (seed 42); across 50 subsamples, the ratio ranges from 1.56 to 1.93 (median 1.75). Sources: \texttt{rq2\_ratio\_sensitivity.csv}, \texttt{rq2\_ratio\_jackknife.csv}, and \texttt{rq2\_ratio\_equal\_n\_draws.csv}, written by \texttt{src/task\_l\_ratio\_sensitivity.py}; the rows are written by \texttt{src/tables\_supplement.py}.}
\label{tab:ratio}
\small
\setlength{\tabcolsep}{3pt}
\begin{tabular}{@{}lrrccccll@{}}
\toprule
& & & \multicolumn{2}{c}{$\tau^2$} & \multicolumn{2}{c}{$\sigma^2$} & & \\
\cmidrule(lr){4-5}\cmidrule(lr){6-7}
Specification & Repos & Agent / human PRs & Agent & Human & Agent & Human & Ratio [95\% CI] & $\Delta\sigma^2$ [95\% CI] \\
\midrule
Primary estimate & 280 & 10{,}623 / 36{,}102 & 1.09 & 0.62 & 2.53 & 3.00 & 1.76 [1.41, 2.18] & [$-$0.78, $-$0.15] \\
Leave one repository out$^a$ & 280 & 10{,}623 / 36{,}102 & -- & -- & -- & -- & 1.71 to 1.80 & -- \\
\midrule
\multicolumn{9}{@{}l}{\itshape Transformation of the outcome} \\
Raw hours & 280 & 10{,}623 / 36{,}102 & 7{,}727 & 4{,}096 & 26{,}226 & 43{,}055 & 1.89 [0.83, 3.75] & [$-$30{,}249, $-$3{,}687] \\
$\log(1+\text{minutes})$ & 280 & 10{,}623 / 36{,}102 & 2.54 & 1.64 & 5.04 & 6.15 & 1.55 [1.26, 1.89] & [$-$1.75, $-$0.46] \\
Rank-based inverse normal & 280 & 10{,}623 / 36{,}102 & 0.28 & 0.19 & 0.57 & 0.69 & 1.47 [1.19, 1.83] & [$-$0.20, $-$0.05] \\
\midrule
\multicolumn{9}{@{}l}{\itshape Calendar time} \\
$M_3$ + creation date & 280 & 10{,}623 / 36{,}102 & 1.06 & 0.57 & 2.53 & 2.98 & 1.84 [1.46, 2.28] & [$-$0.76, $-$0.13] \\
Overlapping creation dates & 183 & 5{,}951 / 14{,}033 & 1.07 & 0.72 & 2.42 & 2.81 & 1.49 [1.14, 2.01] & [$-$0.78, 0.05] \\
\midrule
\multicolumn{9}{@{}l}{\itshape Size of the arms} \\
Both arms capped at 100 & 280 & 9{,}079 / 20{,}602 & 1.09 & 0.64 & 2.57 & 2.86 & 1.72 [1.37, 2.15] & [$-$0.54, $-$0.02] \\
Equal arm sizes$^b$ & 280 & 9{,}097 / 9{,}097 & 1.09 & 0.60 & 2.62 & 2.86 & 1.80 [1.42, 2.31] & [$-$0.54, 0.06] \\
\bottomrule
\end{tabular}
\end{table}

\begin{figure}[!htbp]
\centering
\includegraphics[width=0.5\textwidth]{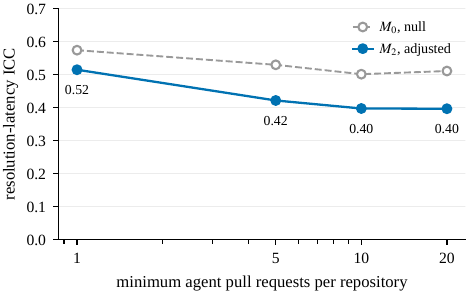}
\caption{Resolution-latency ICC of agent PRs across activity cutoffs, the minimum number of agent PRs a repository must have to enter the sample (1, 5, 10, 20; the paper's primary sample uses 10). Null model $M_0$ and $M_2$, which adds PR, agent, and repository covariates, fitted on each cutoff's resolved PRs (2{,}404, 728, 403, and 205 repositories). Source: \texttt{rq2\_variance\_partition.csv}.}
\label{fig:cutoff}
\end{figure}

\begin{figure}[!htbp]
\centering
\includegraphics[width=0.5\textwidth]{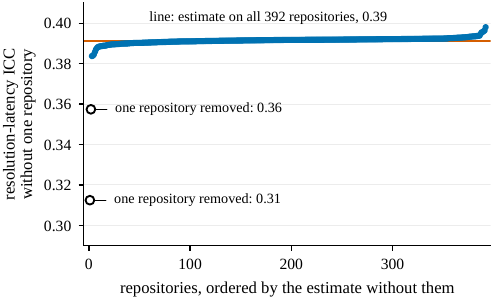}
\caption{Leave-one-repository-out sensitivity of the resolution-latency ICC under $M_3$ on the 392 repositories with API attributes. Each point is the ICC re-estimated without one repository, sorted; the line is the estimate on all 392 (0.39). Removing one repository lowers the estimate to 0.31 and removing another to 0.36; each of the other 390 removals moves it by less than 0.01. Source: \texttt{rq2\_jackknife\_headline.csv}.}
\label{fig:jackknife}
\end{figure}